\documentclass[a4paper,fleqn]{cas-sc}
\usepackage{upgreek}

\usepackage{citesort}
\usepackage[sort&compress, numbers]{natbib}
\usepackage[normalem]{ulem}
\def\tsc#1{\csdef{#1}{\textsc{\lowercase{#1}}\xspace}}
\tsc{WGM}
\tsc{QE}
\tsc{EP}
\tsc{PMS}
\tsc{BEC}
\tsc{DE}

\begin{document}
\let\WriteBookmarks\relax
\def\floatpagepagefraction{1}
\def\textpagefraction{.001}

\shortauthors{J. Doumani et~al.}

\title [mode = title]{Macroscopically Self-Aligned and Chiralized Carbon Nanotubes: From Filtration to Innovation}   
\tnotemark[1]

\tnotetext[1]{This document is the results of the research
   project funded by the Robert A.\ Welch Foundation through Grant No.\ C-1509, the Air Force Office of Scientific Research through Grant No.\ FA9550-22-1-0382, and the Chan Zuckerberg Initiative through Grant No.\ WU-21-357.}

\author[1,2]{ Jacques Doumani}
\ead[URL]{https://profiles.rice.edu/student/jacques-doumani}
\credit{Conceptualization, visualization, writing - original draft, and writing – review and editing}

\author[1]{ Keshav~Zahn}

\author[1,2,3]{ Shengjie~Yu}

\author[1,2]{ Gustavo~Rodriguez~Barrios}

\author[1,6]{ Somesh~Sasmal}

\author[1]{ Rikuta~Kikuchi}

\author[1,2]{ T.~Elijah~Kritzell}

\author[5]{ Hongjing~Xu}

\author[1,4]{ Andrey~Baydin}

\author[1,3,4,5,6]{ Junichiro~Kono}
\ead{kono@rice.edu}
\ead[URL]{https://profiles.rice.edu/faculty/junichiro-kono}

\affiliation[1]{organization={Department of Electrical and Computer Engineering, Rice University},
    addressline={6100 Main Street}, 
    city={Houston},
    postcode={77005}, 
    state={TX},
    country={USA}}

    \affiliation[2]{organization={Applied Physics Graduate Program, Smalley--Curl Institute, Rice University},
    addressline={6100 Main Street}, 
    city={Houston},
    postcode={77005}, 
    state={TX},
    country={USA}}

    \affiliation[3]{organization={Carbon Hub, Rice University},
    addressline={6100 Main Street}, 
    city={Houston},
    postcode={77005}, 
    state={TX},
    country={USA}}
    
    \affiliation[4]{organization={Smalley--Curl Institute, Rice University},
    addressline={6100 Main Street}, 
    city={Houston},
    postcode={77005}, 
    state={TX},
    country={USA}}

    \affiliation[5]{organization={Department of Physics and Astronomy, Rice University},
    addressline={6100 Main Street}, 
    city={Houston},
    postcode={77005}, 
    state={TX},
    country={USA}}

    \affiliation[6]{organization={Department of Materials Science and NanoEngineering, Rice University},
    addressline={6100 Main Street}, 
    city={Houston},
    postcode={77005}, 
    state={TX},
    country={USA}}

\begin{abstract}
Because of their natural one-dimensional (1D) structure combined with intricate chiral variations, carbon nanotubes (CNTs) exhibit various exceptional physical properties, such as ultrahigh electrical and thermal conductivity, exceptional mechanical strength, and chirality-dependent metallicity. These properties make CNTs highly promising for diverse applications, including field-effect transistors, sensors, photodetectors, and thermoelectric devices. While CNTs excel individually at the nanoscale, their 1D and chiral nature can be lost on a macroscopic scale when they are randomly assembled. Therefore, the alignment and organization of CNTs in macroscopic structures is crucial for harnessing their full potential. In this review, we explore recent advancements in understanding CNT alignment mechanisms, improving CNT aligning methods, and demonstrating macroscopically 1D properties of ordered CNT assemblies.  We also focus on a newly discovered class of CNT architectures, combining CNT alignment and twisting mechanisms to create artificial radial and chiral CNT films at wafer scales. Finally, we summarize recent developments related to aligned and chiral CNT films in optoelectronics, highlighting their unique roles in solar cells, thermal emitters, and optical modulators.
\end{abstract}

%
\begin{NoHyper}
\maketitle
\end{NoHyper}
\section{Introduction}

The exploration of carbon nanotubes (CNTs) commenced in the latter half of the 20th century with early studies dating back to the 1950s. Nevertheless, it was in 1991 when Iijima~\cite{iijima_helical_1991} determined the atomic arrangements inside these needle-like tubular structures with nanometer-sized diameters through electron microscopy. 
A single-wall CNT (SWCNT) can be considered a rolled-up version of graphene -- a monolayer of honeycomb-lattice arranged carbon atoms~\cite{graphene}. The way by which the graphene sheet is rolled up to form a tube can be specified by the chiral (or roll-up) vector, denoted by a pair of integers, ($n$,$m$), where $n \ge m$, which classifies SWCNTs into three categories in terms of structure: armchair ($n=m$), zigzag ($m=0$), and chiral ($n \ne m, m \ne 0$)~\cite{Dresselhaus1995}.

Electronically, CNTs come in two fundamental types: metallic and semiconducting~\cite{Dresselhaus2005}. Metallic SWCNTs are gapless in their band structure. Semiconducting species further consist of two sub-categories depending on their band gap, $E_\text{g}$: medium-gap ($E_\text{g}\sim$~1\,eV) and small-gap ($E_\text{g}\sim$~10--100\,meV) semiconducting SWCNTs~\cite{NanotetAl12AM,HarozetAl13NS}. Their diameters (and hence their $E_\text{g}$ values) are influenced by which method of production, such as arc discharge~\cite{arcdisch}, chemical vapor deposition (CVD)~\cite{CVDDD}, and laser ablation~\cite{LaserAblation}, is employed~\cite{Dresselhaus1995}.  In medium-gap SWCNTs, $E_\text{g} \propto 1/d_\text{t}$; in small-gap SWCNTs, $E_\text{g} \propto 1/d^2_\text{t}$, where $d_\text{t}$ is the tube diameter.

As a result of their unparalleled thermal, optical, mechanical, and electrical properties~\cite{balandin2011thermal,tomanek2008introduction,mechanical_review,lekawa2014electrical,WeismanKono19Book}, CNTs quickly captivated the attention of both academics and industrial scientists. These properties are promising for their use in diverse applications: their ultrahigh electrical conductivity and strong mechanical structure allow for use as power-grid wire materials~\cite{behabtu2013strong,WangetAl14AFM,Lee2022cond}; their ultrahigh thermal conductivity and high thermoelectric power factors allow for use for on-chip thermal management~\cite{komatsu2021macroscopic}; their high optical anisotropy allows for use as polarizers~\cite{RenetAl09NL}, waveplates~\cite{BaydinetAl21Optica}, polarization-sensitive detectors~\cite{NanotetAl13SR,HeetAl14NL}, emitters of polarized thermal radiation~\cite{MatanoACS2023,matano2022electrical,GaoEtAl2019ACSP}, and full-duplex directional beam-forming systems~\cite{10298902}; their high on/off ratios allow for use in field-effect transistors (FET)~\cite{fieldeffect,kang2007high,avouris2007carbon}; and their power of tunable single-photon emission allows for use as room-temperature quantum emitters in the telecommunication band~\cite{Ma2015,Ishii2018}.

The utilization of these exceptional properties is contingent on a single factor: the axial alignment of CNTs~\cite{lan2011physics}. If CNTs are randomly assembled, any physical property relying on the quantum confinement and 1D character of carriers and phonons in individual CNTs is averaged out across all directions, resulting in the loss of intrinsic 1D properties in macroscopic CNT structures. For these superb properties to be present at a macroscopic scale, CNTs must be globally aligned when assembled. Therefore, developing a repeatable, scalable, and economical process of ordering CNTs into macroscopic structures is crucial.

During the last two decades, various techniques have emerged for aligning CNTs. Alignment can occur \textit{in situ}, i.e., during the growth process, under certain conditions, as demonstrated in instances such as directional gas flow and electric field-assisted CVD~\cite{zhang2001electric,HuangEtAl2003JPC,murakami2004growth,ding2008growth,pint2008formation}. Alternatively, post-growth alignment can be achieved starting from CNT suspensions through methods such as vacuum filtration~\cite{deHeeretAl95Science,Shaffer1998,DanetAl12IECR}, wet spinning~\cite{vigolo2000macroscopic,ericson2004macroscopic,behabtu2013strong}, mechanical stretching~\cite{kim2005highly,fagan2007dielectric,guo2022soft}, shear-force~\cite{kim2008capillarity,jinkins2019substrate,liu2020aligned}, the Langmuir--Blodgett method~\cite{li2007langmuir,joo2014dose}, the Langmuir--Schaefer method~\cite{cao2013arrays}, dielectrophoresis by applying an AC electric field~\cite{DimakiEtAl2005N,PadmarajEtAl2008N,shekhar2011ultrahigh}, and magnetic alignment~\cite{SmithEtAl2000APL,Walters2001}. 
Each method offers unique advantages suited for specific applications. The recently developed controlled vacuum filtration (CVF) method~\cite{He2016,gao2019science}, for example, can create highly aligned CNTs with packing densities higher than any other methods, aligning any CNT species at the wafer scale while reaching nematic orders of nearly 1, without requiring expensive tools. 

In this review, we focus on recent developments in CNT alignment using the CVF method and their applications. We first examine the significance of filter membrane grooves in CVF and their impact on alignment. We discuss image analysis techniques for alignment characterization and quantification, and further explore two new types of CNT architectures: radial alignment and chiral assemblies. We also cover CVF automation and its ability to produce multiple aligned films concurrently. Additionally, we delve into the optical properties of aligned and chiral films under linearly polarized light and circularly polarized light (CPL), respectively. Lastly, we analyze the practical applications of CVF-derived aligned CNT films, including THz wave manipulation and solar cell performance enhancement.

\section{Fabrication of Macroscopically Ordered CNT Assemblies}
The advancement of the frontiers of ordered CNT research requires a comprehensive understanding of alignment mechanisms. This knowledge is the foundation for optimizing fabrication processes, the precise local and global characterization of nematic order, the enhancement of automation to bolster reproducibility and scalability, and the innovation of novel structures driven by cutting-edge scientific insights.

\begin{figure}
    \centering
    \includegraphics[width=\textwidth]{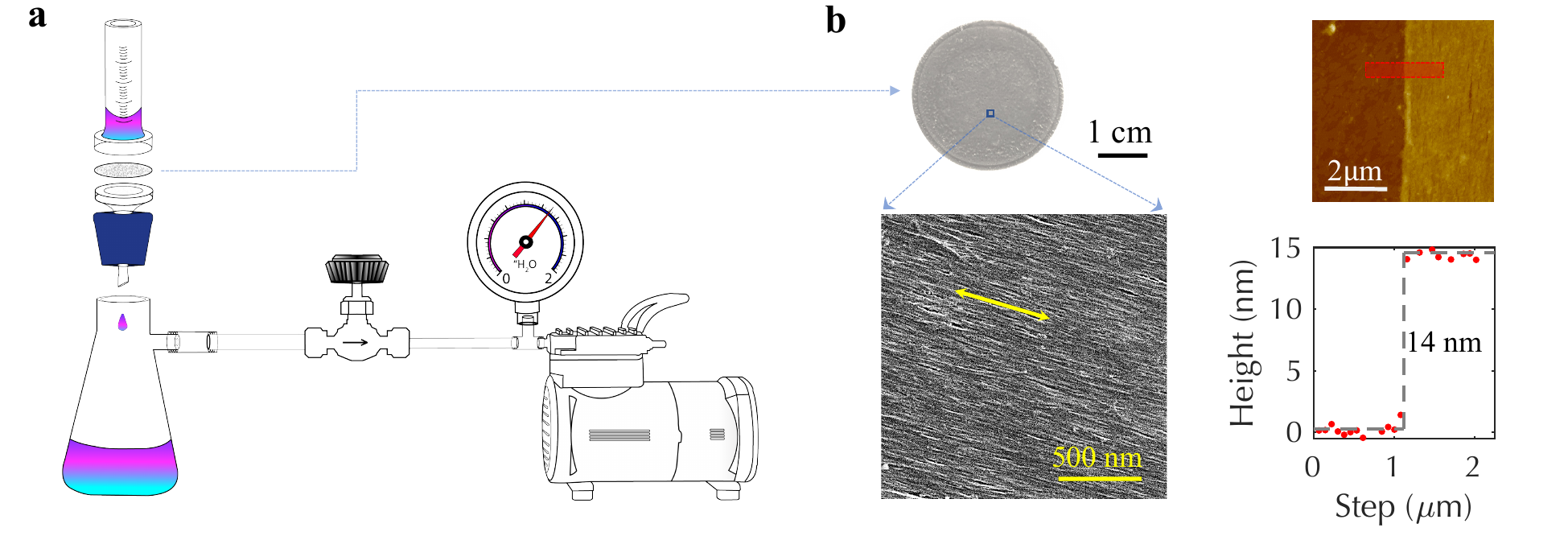}
    \caption{Fabricating aligned CNT films using the controlled vacuum filtration (CVF) method. (a)~Illustration of the vacuum filtration setup. (b)~Visual representation (top left), SEM image (bottom left), AFM image (top right), and corresponding height profile (bottom right) of a CVF-produced aligned CNT film. Adapted from~\cite{doumani_controlled_2023}. \href{https://creativecommons.org/licenses/by/4.0/}{CC BY 4.0.}} 
    \label{figure_1}
\end{figure}

\begin{figure}[t!]
    \centering
    \includegraphics[width=\textwidth]{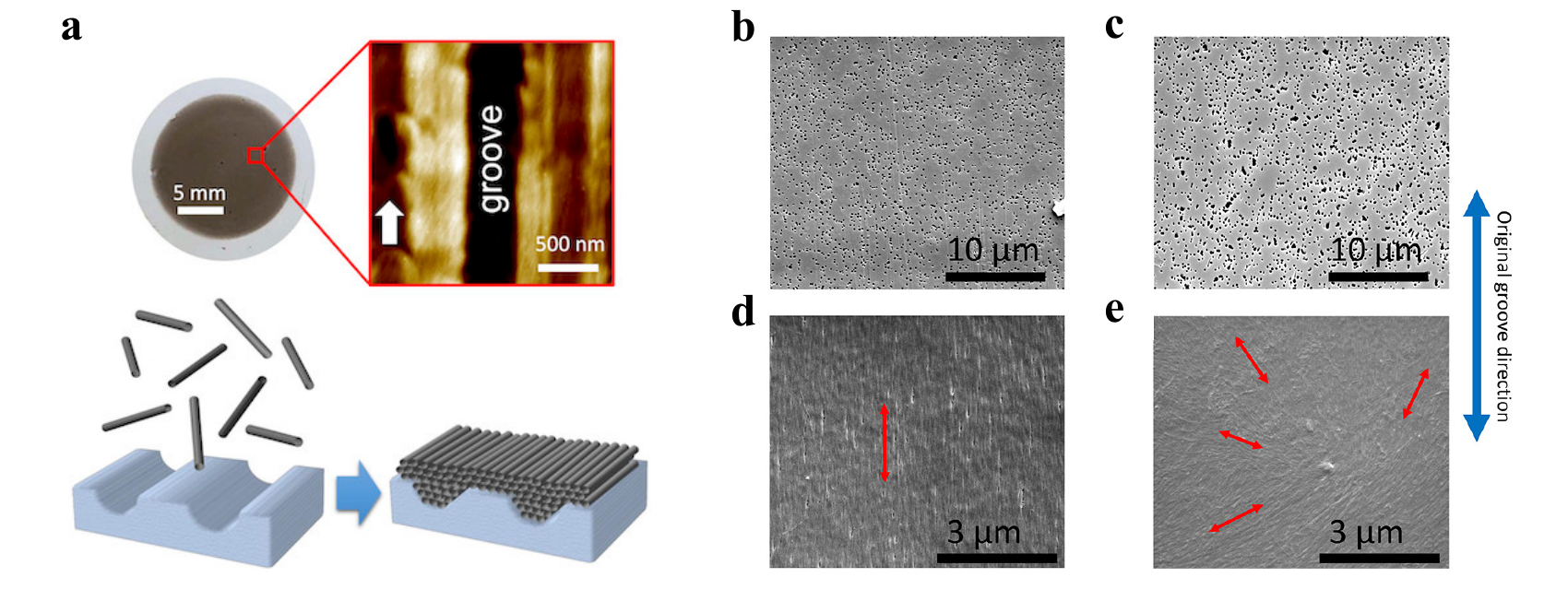}
    \caption{Effect of grooves imprinting on the global alignment in CVF. (a)~Photograph of a wafer-scale film (inset: AFM image showing a micro-groove) (top), schematic of grooves/nanotube assembly before and after filtration (bottom). SEM images of (b)~untreated and (c)~treated filter membrane. (d)~SEM image of a CNT film produced on the untreated membrane, displaying an alignment director along the groove direction (arrow). (e)~SEM image of a CNT film produced on the heated membrane, showing no global alignment but domains of aligned CNTs with random directions (arrows). The blue arrow indicates the original groove direction before heating in all images. Adapted with permission from~\cite{Komatsu2020}. Copyright 2020 American Chemical Society.}
    \label{figure_2}
\end{figure}

\subsection{The controlled vacuum filtration method}
Vacuum filtration has long been used to fabricate CNT films, and some early studies (e.g.,~\cite{deHeeretAl95Science,Shaffer1998,DanetAl12IECR}) have noted spontaneous alignment of CNTs. In 2016, He \emph{et al}.~\cite{He2016} developed the \emph{controlled} vacuum filtration, or CVF, method for preparing highly aligned, densely packed, single and mixed chirality, and wafer-scale mono-domain CNT films. This technique was made possible through control of surfactant and CNT concentrations, filter membrane treatment, filtration parameters, and CNT fabrication processes.

\subsubsection{Basic procedures and components}
The CVF process involves the filtration of an aqueous solution of surfactant-suspended CNTs through a filter membrane~\cite{He2016,gao2019science}. As depicted in Fig.~\ref{figure_1}(a), a basic CVF apparatus comprises a funnel, a thin polycarbonate film with micrometer-scale pores, a fritted glass support, and a vacuum flask. An outward flow is enabled by connecting the main filtration assembly with a vacuum pump, finely regulated with a valve and vacuum gauge.

This filtration selectively retains the CNTs, forming a film on the membrane's surface. To achieve global alignment, specific concentration thresholds for both surfactants and CNTs in the suspension are crucial, and the vacuum filtration process must be slow and well controlled. Notably, He \emph{et al.}'s approach~\cite{He2016,gao2019science} enables the production of uniform, wafer-scale SWCNT films, as can be seen in Fig.~\ref{figure_1}(b), with adjustable thickness ranging from a few to hundreds of nanometers. CVF-produced films display an ultrahigh cross-sectional areal density of approximately $1 \times 10^6$\,CNT $/\upmu$m$^{2}$ and exhibits a minimal angular standard deviation in CNT orientation, measuring only $1.5^\circ$, highlighting the capacity of CVF to produce highly aligned CNT films. Notable anisotropy in absorption is observed between light polarized parallel and perpendicular to the alignment direction (see Sec.~\ref{spectro}).

Moreover, the CVF approach can be seamlessly integrated with established microfabrication techniques that are commonly employed in the fabrication of various electronic and optoelectronic devices~\cite{wang2020ultrahigh, matano2022electrical, GaoEtAl2019ACSP, RobertsEtAl2019NL}. Furthermore, it is worth noting that the CVF method applies to CNT ensembles of both mixed and purified chirality, i.e., it has the capability of aligning single-chirality-enriched CNT species, such as (6,5) SWCNTs.

The alignment mechanism proposed by He \emph{et al.}~\cite{He2016} relies on two-dimensional surface confinement. A hydrophilic coating imparts a negative charge to the filter membrane surface during the filtration process. Consequently, negatively charged CNTs experience repulsion from the membrane surface. Simultaneously, van der Waals forces existing between the CNTs and uncoated regions of the membrane facilitate the accumulation and formation of an ordered two-dimensional phase of aligned CNTs. This mechanism, unlike traditional three-dimensional nematic ordering, allows for accumulation of well-aligned CNTs into thick films.

More recently, Komatsu~\emph{et al}.~\cite{Komatsu2020} have provided insightful information on the alignment mechanism underlying the CVF method, specifically clarifying what dictates the alignment direction. They discovered that polycarbonate track-etched filter membranes featured parallel ``macrogrooves'' that were comprised of collections of smaller ``microgrooves''; see Fig.~\ref{figure_2}(a). These groove dimensions varied between batches, and, crucially, the authors found that these grooves were the determiners of alignment with the direction of these surface-imprinted grooves matching the direction of alignment; see Figs.~\ref{figure_2}(b) and (d). 

Furthermore, Komatsu~\emph{et al}.~\cite{Komatsu2020} devised a novel method to eliminate these grooves through a heat treatment process, in which a filter membrane was heated beyond its glass transition temperature and compressed, removing the surface grooves and, thereby, alignment; see Figs.~\ref{figure_2}(c) and (e).
They demonstrated the ability to controllably re-pattern new grooves on the filter membrane using a diffraction grating with a specific groove density. This re-patterning technique allowed them to produce globally aligned CNT films with a controllable CNT alignment direction through CVF with filter membranes featuring artificial grooves. These experiments showcased the controllability of the CNT alignment direction by manipulating the presence and orientation of grooves on the filter membrane, shedding light on a critical aspect of the alignment process.

\subsubsection{Alignment characterization through SEM image processing}

\begin{figure}[t!]
    \centering
    \includegraphics[width=\textwidth]{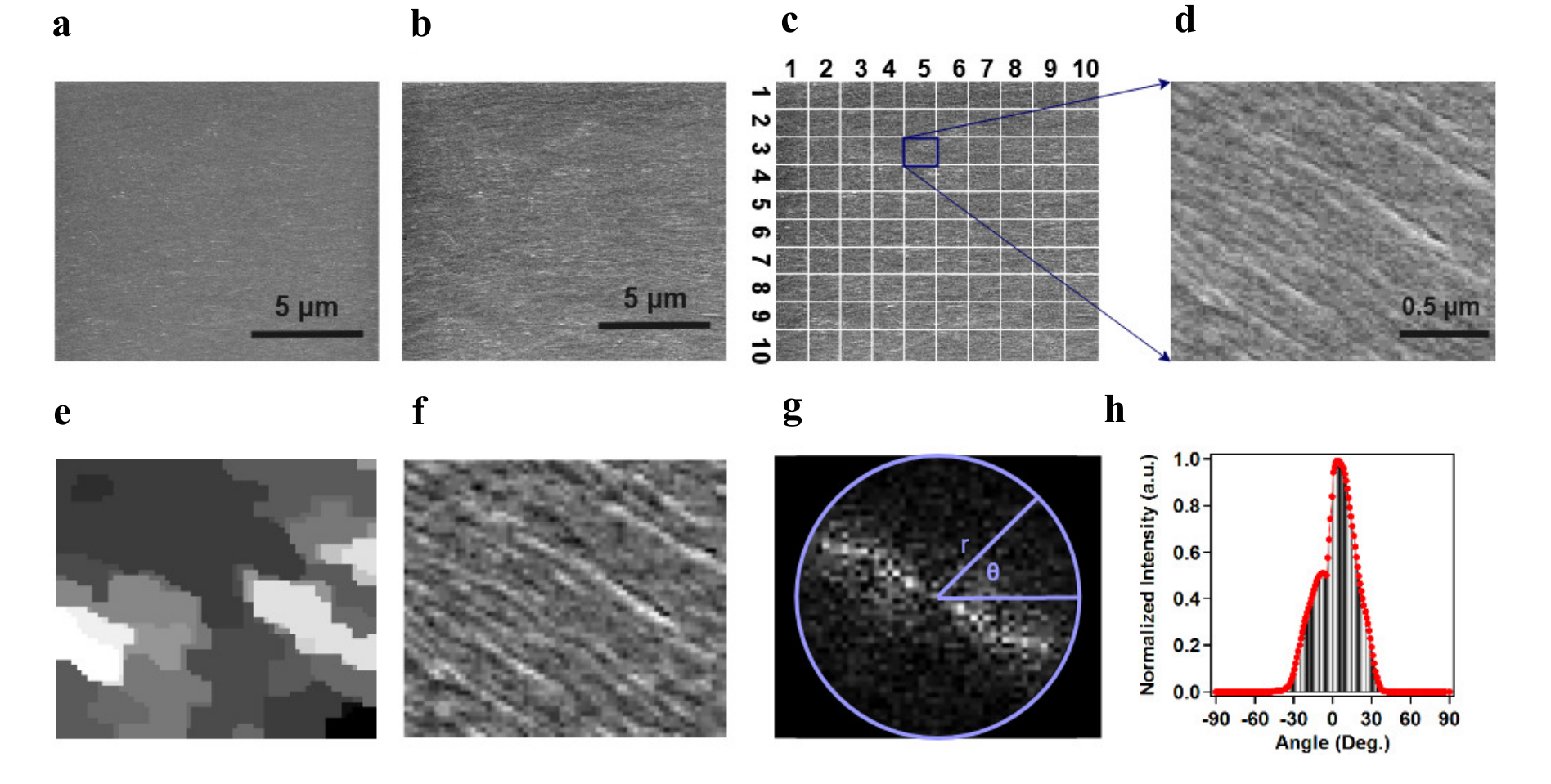}
    \caption{Determining the degree of alignment in CNT films through SEM image processing: (a)~Original SEM image of the sample. (b)~Enhanced image using Laplacian filtering for edge detection. (c)~Segmentation of image (b) into 100 uniform blocks. (d)~Randomly selected block from (c). (e)~Structural component of (d) obtained through decomposition. (f)~Textural component of (d) obtained through decomposition. (g)~Radial summation of the 2D-FFT power spectrum of (f). (h)~Integrated orientation distribution derived from combining individual orientation distributions of all image blocks. Reprinted from \cite{Imtiaz2023}, Copyright (2023), with permission from Elsevier.}
    \label{figure_7}
\end{figure}

In the process of optimizing the fabrication of aligned CNT films, the characterization of nematic order (denoted by $S$) is pivotal. In this context, $S=1$ signifies perfect alignment while $S=0$ represents complete misalignment. Traditional methods for assessing CNT alignment, such as polarized Raman spectroscopy~\cite{Zamora2008,Liu2014,Wang2006}, polarized photoluminescence (PL) excitation (PLE) spectroscopy~\cite{Miyauchi2006, Lefebvre2007}, and polarized optical absorption spectroscopy~\cite{KatsutaniEtAl2019PR,zubair2016carbon,He2016,RenetAl09NL}, while effective, come with complexities, time-consuming procedures, and dependence upon knowledge about the CNT species contained in the sample. This complicates the analysis process, especially with samples containing multiple chiralities. 
Additionally, more advanced methods, such as small- and wide-angle X-ray scattering~\cite{Meshot2010,Wei2002}, offer valuable insights into alignment but are often impractical for large-scale analysis due to their resource-intensive nature. 
Therefore, the goal is to develop a robust and universally applicable method for the characterization of nematic order for a variety of CNT samples.

Imtiaz \emph{et al}.~\cite{Imtiaz2023} have recently introduced a scanning electron microscopy (SEM)-based image processing technique for characterizing CNT alignment and determining the nematic parameter, $S$. This method involves the acquisition of SEM images at various sample locations, as exemplified in Fig.~\ref{figure_7}(a). Subsequent steps in the image processing pipeline, shown in Figs.~\ref{figure_7}(b)-(f), include Laplacian filtration for edge enhancement, block partitioning, and decomposition into structural and textural components. The orientation distribution is determined through the 2D-FFT power spectra of each block, culminating in the computation of a global orientation across multiple blocks; see Figs.~\ref{figure_7}(g)-(h). Notably, the technique can detect and address cracks and holes through a thresholding approach, effectively converting them into high-intensity regions. This feature enhances the method's versatility, making it suitable for various morphological variations. Importantly, this technique displays promising comparisons with terahertz (THz) time-domain spectroscopy, indicating a potentially more accessible and efficient means to assess CNT alignment and nematic order.

The nematic order parameter $S$ in two dimensions is calculated by
\begin{align}
    S_\text{2D}=\frac{ \int_{0}^{\pi} I(\theta)(2\cos^{2}\theta -1) \,d\theta}{ \int_{0}^{\pi} I(\theta) \,d\theta},
    \label{SEM}
\end{align} 
where $I(\theta)= (1/k) \sum_i {I_i (\theta)}$, $I_i (r,\theta) = \sum_s{F_s (r,\theta)}$ is the orientation distribution, and $F_s (r,\theta)$ is the polar component of the power spectrum for a particular radius $r$ and angle $\theta$.

\subsection{Novel CNT architectures}\label{CD_fab}
The propagation of electrons, photons, and phonons through matter is profoundly influenced by the structural characteristics of the matter. Many previous studies of aligned CNT systems have shown strongly anisotropic electrical, optical, and thermal properties at a macroscopic scale~\cite{gao2019science}.  Recent studies have shown that the CVF method can be used not only for fabricating macroscopically aligned films but also for creating macroscopically order CNT assemblies With different types of broken symmetries.  In this section, we describe two examples of fabrication of such novel assemblies: CNT films with synthetic chirality and CNT films with radial alignment.

\begin{figure}[t!]
    \centering
    \includegraphics[width=\textwidth]{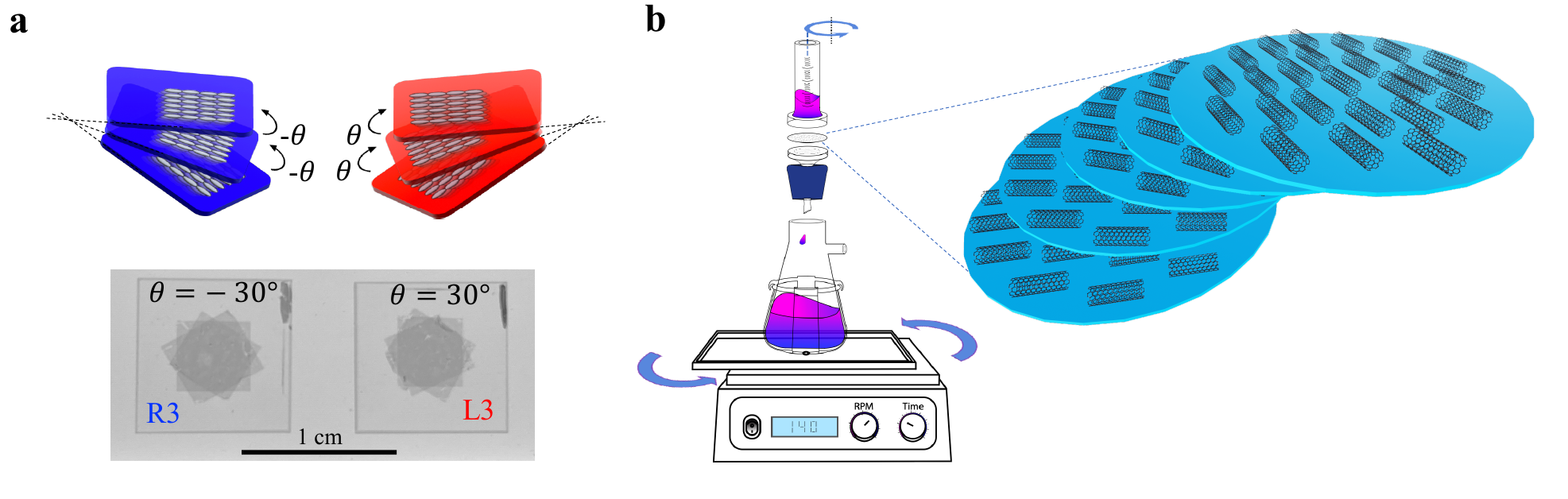}
    \caption{Chiralized CNT film formation using the CVF method. (a)~Twist-stacked CNT films -- schematics and photographs of CVF-produced aligned CNT films twist-stacked in right-handed and left-handed manners. The twist angle was 30 degrees. (b)~Mechanical-rotation-assisted CVF (orbital shaking) -- a standard filtration system mounted on an orbital mechanical shaker, resulting in a twisted CNT thin-film architecture. Adapted from~\cite{doumani_controlled_2023}. \href{https://creativecommons.org/licenses/by/4.0/}{CC BY 4.0.}} 
    \label{figure_3}
\end{figure}

\subsubsection{Chiral assemblies}

Chirality -- an intrinsic characteristic of specific molecules and structures -- determines enantiomers, i.e., two mirror-image forms.  Despite their identical chemical and mechanical properties, the two enantiomers exhibit distinct behaviors in response to circularly polarized light, leading to chiroptical phenomena.  These phenomena are responsible for the complex patterns of seashells, the asymmetry of amino acids, and the complex structures of biomolecules, such as proteins and DNA~\cite{liu_supramolecular_2015}.

The most commonly studied chirality exists at the molecular level, but chirality can also be artificially synthesized in macroscopic systems, where ordered structures with twisting patterns play a crucial role. Such patterns can be created through various techniques, including the additive manufacturing of polymer-nanoparticle composites~\cite{yang_biomimetic_2017}, temperature-controlled assembling~\cite{schulz_giant_2018}, self-assembly processes~\cite{albano_chiroptical_2020}, nanofabrication of metamaterials~\cite{wang_optical_2016}, and mechanical deformation~\cite{kim_reconfigurable_2016}.

In the context of CNTs, abundant evidence exists for the molecular-level chirality arising from the handed arrangements of carbon atoms, when the chirality indices $n$ and $m$ differ and $m \ne 0$. However, structural chirality in macroscopic CNT assemblies have remained unexplored.
Very recently, building upon the CVF work of He \emph{et al}.~\cite{He2016}, Gao \textit{et al}.~\cite{gao2019science}, and Komatsu \emph{et al}.~\cite{Komatsu2020}, Doumani \emph{et al}.~\cite{doumani_controlled_2023} have developed two methods for large-scale production of artificial chiral CNT films. In the first, \textit{ex~situ}  method, they stacked highly aligned CNT films onto a substrate, where subsequent layers underwent controlled 3D twisting to achieve specific angles upon contact; see Fig.~\ref{figure_3}(a). This approach enabled continuous `AB'-, `ABC'-, and alternating `ABAB'-type stacking, with a 14-nm-thickhighly aligned CNT film building block. They demonstrated that, by controlling the twisting handedness, it is possible to precisely dictate the chirality; see Sec.~\ref{spectro} for more details.

The second, \textit{in~situ} method of Doumani \emph{et al}.~\cite{doumani_controlled_2023} introduced twist-orbital shaking during CVF; see Fig.~\ref{figure_3}(b). In this approach, the CVF system is mounted on an orbital shaker. By introducing brief bursts of shaking during the controlled filtration process, while controlling the movement amplitude, period, speed, and deceleration, the authors were able to induce alignment and swirling effects, which resulted in synthetic chiral structures. This method offers a one-shot chirality engineering technique without convoluted procedures, opening new avenues for fabricating chiral assemblies in 1D and 2D materials. In Sec.~\ref{spectro}, we describe the chiroptical response of these chiral films using circular dichroism (CD) spectroscopy.

\subsubsection{Radial alignment of CNT}

\begin{figure}[t!]
    \centering
    \includegraphics[width=\textwidth]{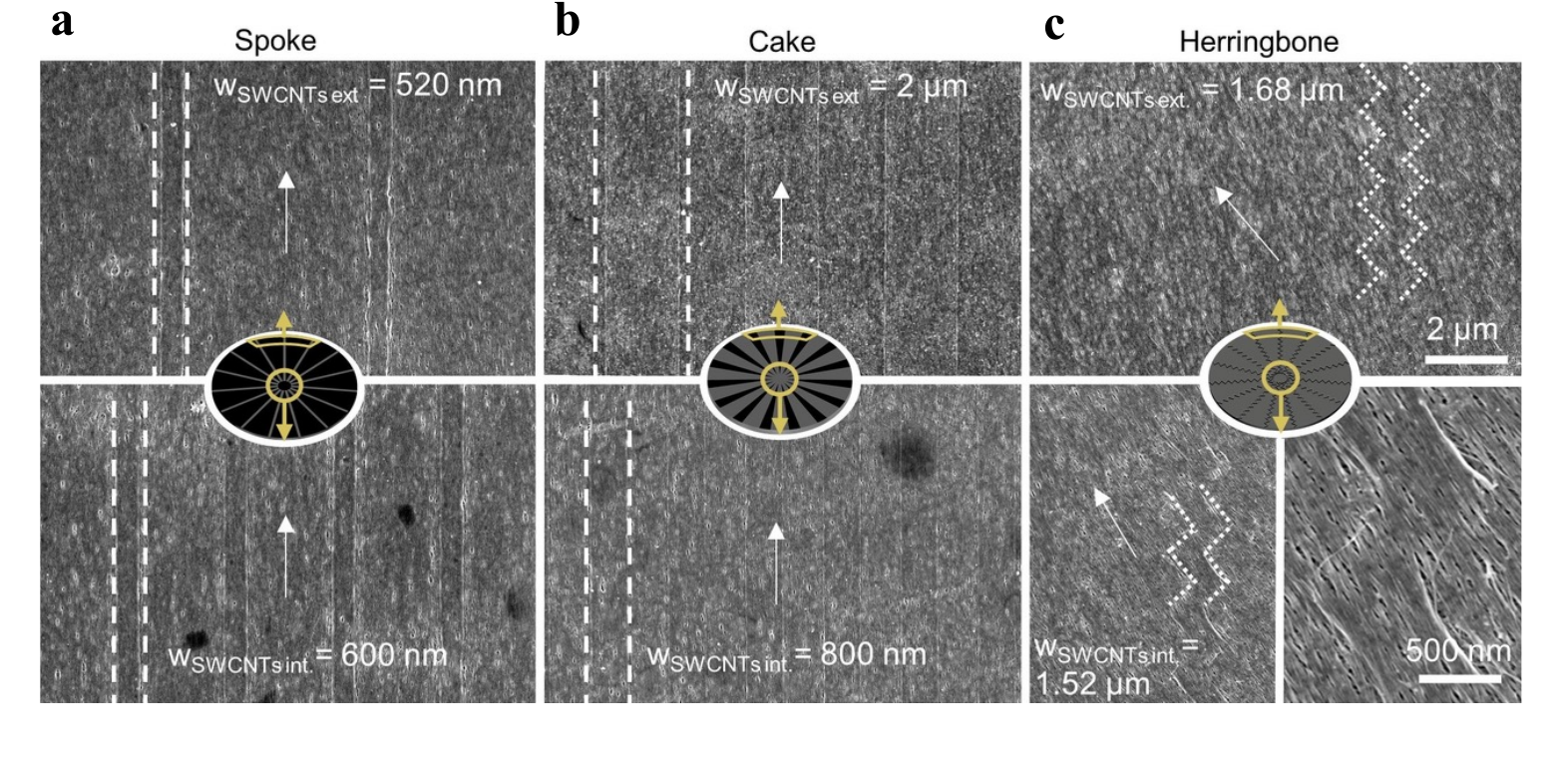}
    \caption{SEM images of SWCNT films on polyamide flexible printed circuit boards created from membranes of various shapes: (a)~spoke (SP), (b)~cake (CA), and (c)~herringbone (HB) patterns. Arrows indicate the alignment director for clarity. Adapted from~\cite{Rust2023}. \copyright\,~2023 The Authors. Small published by Wiley-VCH GmbH. \href{https://creativecommons.org/licenses/by/4.0/}{CC BY 4.0.}} 
    \label{figure_4}
\end{figure}

In a recent study, Rust \emph{et al}.~\cite{Rust2023} have explored the radial alignment of SWCNTs by employing novel fabrication processes. They used electron beam lithography to create precision polymer masks with `spoke,' `cake,' and `herringbone' patterns, as depicted in Figs.~\ref{figure_4}(a-c), respectively, enabling hot-embossing of polycarbonate track-etched filter membranes for the production of radially aligned CNT films.

The imprinted membranes were integrated into the dead-end filtration system, analogous to CVF, detailed in Sec.~\ref{autom}. After filtering out the CNT solution, these films were transferred onto solid substrates or printed circuit boards that are fleixble for further examination. The resultant radial alignment of SWCNTs replicated the mask patterns successfully, resulting in films with consistent radial orientation, confirmed through polarized spectroscopy techniques. Rust \emph{et al}.\ postulated that this alignment is predominantly driven by a collective shear flow, instead of electrosatic forces. In their observations, they highlighted that under cross-polarized light, the assemblies of CNTs displayed a unique flower-like light pattern characterized by distinct dark and bright regions. Notably, this pattern remained unchanged regardless of the sample orientation. In the concluding phase, a two-probe resistance measurement revealed a one-third reduction in resistance for CA radial films compared to a random films.

Beyond the confines of the laboratory, radially aligned SWCNT films hold promise for practical applications. Notably, they can be used in the field of optics, serving as radial polarizers~\cite{chen2007carbon} and photomasks in the extreme ultraviolet wavelength range~\cite{gubarev2019single}. Furthermore, their potential extends into fields such as thermal management, where they can contribute to efficient and uniform heat distributions~\cite{yu2021advances}.

\subsection{Automation of controlled vacuum filtration}\label{autom}
Amidst all the advancements we discussed, optimized reproducibility and consistency have remained underexplored. Fortunately, these elements have not been overlooked. The CVF technique is dependent on the maintenance of a consistent filtration rate through a filter membrane to attain a high degree of alignment. However, ensuring this steady filtration rate manually can be a challenge, prone to variability and human errors that may compromise the quality of the resultant films. To address this issue, a solution has emerged in the form of automation integrated into the filtration process. Automation plays an important role in securing the uniform alignment of SWCNTs after each filtration. This forward step ensures not only the reproducibility of results but also the much-needed consistency in the production of high-quality films.

 \begin{figure}[t!]
    \centering
    \includegraphics[width=\textwidth]{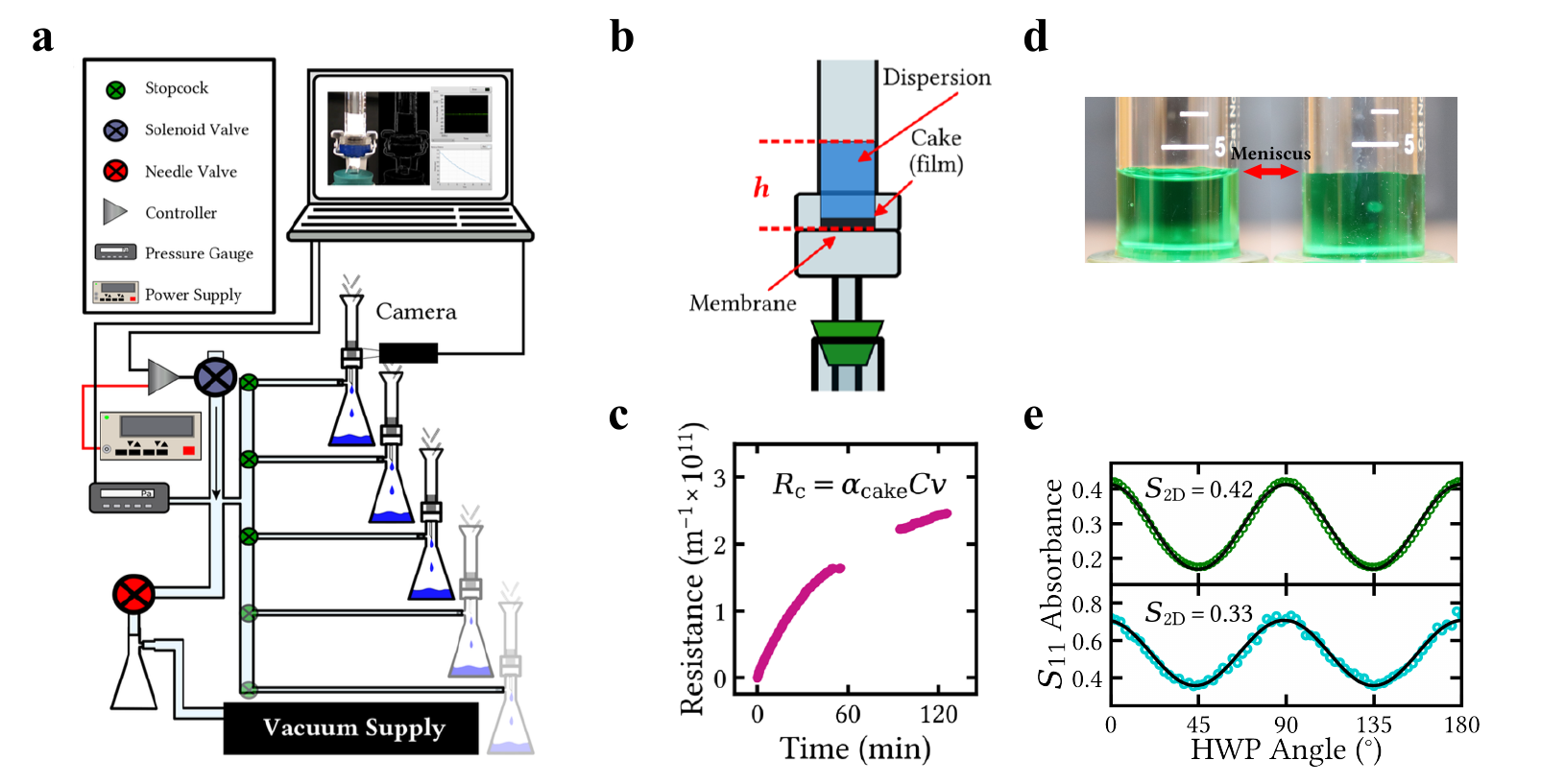}
    \caption{Automating the CVF. (a)~Illustration of the automated and parallelized CVF system. (b)~A closer look at the funnel-membrane-fritted glass support assembly. The height $h$ is tracked off in a PID feedback loop. (c)~Experimental cake resistance as a function of filtration period. (d)~Image of the funnel before (left) and after (right) silane treatment with (e)~the associated 2D nematic order parameter characterized at the $E_{11}$ interband transition of semiconducting SWCNTs (bottom) before and (top) after silane treatment. Adapted with permission from~\cite{WalkerEtAl2019NL}. Copyright 2019 American Chemical Society.}
    \label{figure_5}
\end{figure}

\subsubsection{Conventional vacuum filtration automation and parallelism}
To further advance CNT alignment using the CVF method, Walker \emph{et al}.~\cite{WalkerEtAl2019NL} have automated their system with flow control through a solenoid valve, camera, electric pressure gauge and other electronic components, such that a programmed system tracks and controls a SWCNT meniscus during the filtration process; see Fig.~\ref{figure_5}(a). Their software used a self learning algorithm to convert images of the filtration setup into an edge outline, which represented the meniscus of the SWCNT solution with a parameter $h$, as illustrated in Fig.~\ref{figure_5}(b), allowing for the determination of the solution flow rate.

A feedback system, which used the data obtained from the computer vision system based on the proportional-integral-derivative (PID) control scheme, controlled the flow rate by regulating the vacuum (negative) pressure when gravity pushed down on the solution. Pressure control was achieved using a PID-controlled variable leak through a proportioning solenoid valve. The authors further demonstrated the possibility of CVF-parallelism, that is, the production of multiple films simultaneously through parallel and automated filtration. Although alignment was limited to a lower nematic order parameter, all films showed almost identical degrees of alignment and thicknesses, allowing for CVF to be scaled up. 

Along with automation and parallelism, Walker \emph{et al}.~\cite{WalkerEtAl2019NL} empirically determined the membrane resistance ($R_\text{m}$) and cake resistance ($R_\text{c}$), shown in Fig.~\ref{figure_5}(c), to achieve a constant flow rate ($J$): $1/J=\mu(R_\text{m} + R_\text{c})/\Delta P$, where $\mu$ is the viscosity of the permeate (SWCNT solution), $\Delta P$ is the transmembrane pressure or summation of the applied and head pressure $\rho gh(t)$, where $\rho$ is the dispersion mass density, $g$ is the gravitational acceleration, and $h(t)$ is the solution column height. 

Finally, as shown in Figs.~\ref{figure_5}(d)-(e), the authors reported that the degree of alignment could be enhanced by performing a silane treatment of the funnel; that is, in the glass surface preparation process, acetone and water were used for cleaning, followed by chemical etching using a buffered oxide etchant. Afterwards, a glass steam treatment was performed to incorporate more water into the glass. Finally, a silane reaction occurred in a vacuum desiccator with dimethyldichlorosilane solution. This process resulted in a noticeable change in the meniscus shape in the glass funnel before and after silanization.

\begin{figure}[t!]
    \centering
    \includegraphics[width=\textwidth]{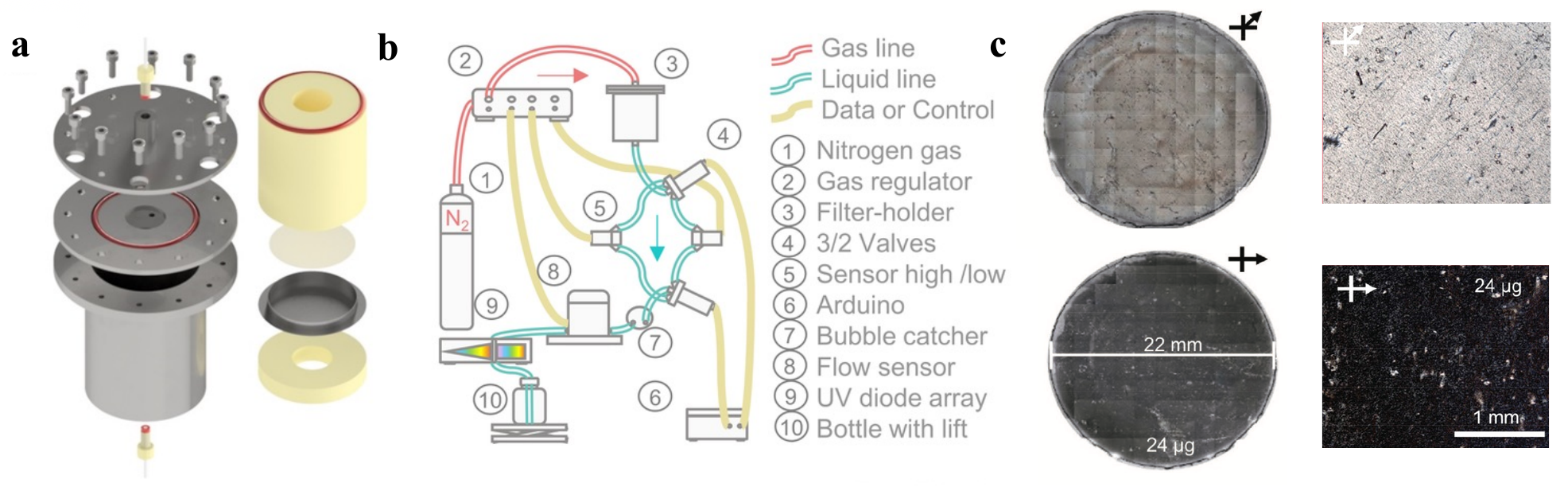}
6,     \caption{Dead end filtration. (a)~Filter holder assembly. (b)~A schematic diagram of the filtration setup. (c)~Combined cross-polarized microscopy images of an aligned CNT film produced using the dead-end filtration setup, fabricated with 24~$\upmu$g of CNTs. Inset: cross-polarized optical images of one spot on the aligned film. Adapted from~\cite{RustEtAl2021AFM}. \copyright\,2021 The Authors. Advanced Functional Materials published by Wiley-VCH GmbH. \href{https://creativecommons.org/licenses/by/4.0/}{CC BY 4.0.}}
    \label{figure_6}
\end{figure}

\subsubsection{Dead end filtration}
While the CVF method relies on a negative-pressure drive, Rust \emph{et al}.~\cite{RustEtAl2021AFM} have developed a new approach to achieve the global alignment of CNTs with a custom microfluidic positive-pressure cell, further advancing Walker \emph{et al}.'s automation process. Previous methods used glass or metallic frit-supported membranes with a vacuum pump, but they lacked inline measurement capabilities and precise initial conditions. The newly designed microfluidic cell, shown in Fig.~\ref{figure_6}(a), addressed these issues and allowed for accurate control of flow rates from a zero-flow state. The microfluidic cell consisted of a membrane supported by a stainless-steel mesh with Teflon blocks on either side, held in a stainless-steel sleeve by flanges. Positive pressure was applied through a channel of pressured nitrogen gas regulated by a digital pressure regulator. Figure~\ref{figure_6}(b) shows the fully designed system, where the permeate was controlled by solenoid valves and measured using a high- and low-pressure piezoelectric sensor, a Coriolis flow sensor, and a photodetector; the permeate was then collected in a waste container.

By calibrating the pressure sensors for precise flow control with a feedback loop from the Coriolis sensor, Rust \emph{et al}.~\cite{RustEtAl2021AFM} characterized the filtration process as described by the Darcy law, $R_\text{total} = p_\text{TMP}/\mu J$, where $\mu$ is the viscosity of the solvent, $J$ is the fluid flux quantified by $\dot{v}=dV/dt$ normalized by the membrane area $A_\text{m}$, and $p_\text{TMP}$ is the direct inline transmembrane pressure $p_\text{in} - p_\text{out}$.
The authors described a series of accumulating resistances over time during filtration,  $R_\text{total} = R_\text{setup} + R_\text{m} + R_\text{cp} + R_\text{b}$. Here, $R_\text{setup}$'s contribution is negligible compared to the membrane resistance ($R_\text{m}$), the resistance due to concentration polarization ($R_\text{cp}$), and the irreversible fouling ($R_\text{b}$). 

The authors identified various mechanisms for SWCNT accumulation at the membrane surface by fitting a power law to $R_\text{b}$. The mechanisms are described as complete blocking, intermediate blocking, standard blocking, and cake filtration. In Fig.~\ref{figure_6}(c), cross-polarized optical images of an aligned CNT film obtained through the dead-end filtration technique are presented. The homogeneous and pronounced contrast observed between the two polarization configurations serves as a clear indicator of the globally achieved high degree of alignment within this specific setup. The setup and methodology described in the research provided an accurate and controlled method for globally aligning SWCNTs. The custom microfluidic cell allows for real-time measurements and offers insights into the filtration process and SWCNT alignment mechanisms.

\section{Optical Spectroscopy of Macroscopically Ordered CNT Assemblies} \label{spectro}

Room-temperature-operating quantum light sources are needed, akin to the core logic unit in quantum computing and to transistors in classical binary computing. These light sources must exhibit robustness due to quantum confinement, enabling emissions in the telecommunication range, while also being easy to control and environmentally friendly. CNTs present a promising candidate for this role~\cite{HeEtAl2017NP,baydin2022carbon,HeEtAl2016NN,heller2020banning}, as they harness profound quantum confinement due to their ultrasmall tube diameters. These dimensions, ranging from 1 to tens of nanometers, are accompanied by the presence of subbands and van Hove singularities in their density of states~\cite{kim1999electronic,NanotetAl13Book}, resulting in strongly excitonic spectroscopic features through interband transitions within the telecommunication range~\cite{ostojic2004interband,NanotetAl12AM,WeismanKono19Book}. Studying the optical properties of aligned and chiral CNTs architectures is crucial, as it paves the way for achieving scalable quantum light sources on integrated photonic chipsets.

\begin{figure}[t!]
    \centering
    \includegraphics[width=\textwidth]{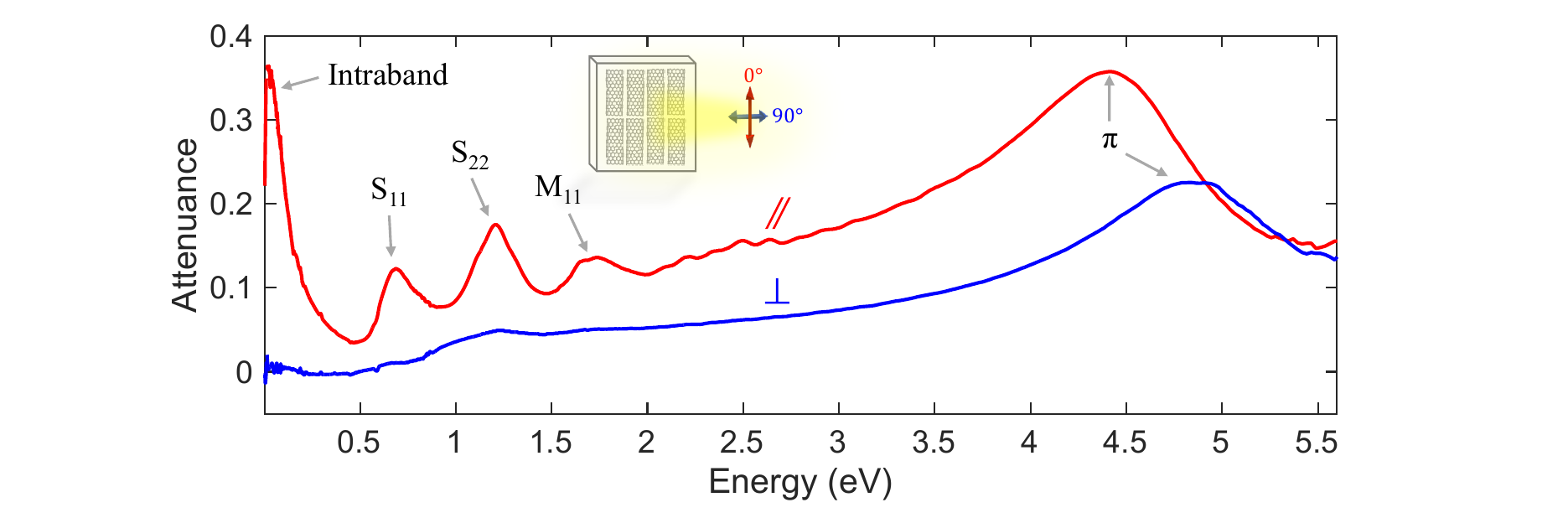}
    \caption{Optical attenuation spectra for a highly aligned CNT film produced by using the CVF method, showing strong polarization dependence, spanning from the terahertz (THz) to the deep ultraviolet (DUV) spectral range~\cite{He2016,doumani_controlled_2023}.}
    \label{figure_7_new}
\end{figure}

\subsection{Linear polarization spectroscopy and hyperbolic properties}
Highly aligned CNT films, produced using the CVF technique, exhibit strong optical anisotropy (Fig.~\ref{figure_7_new}). For perpendicular polarization, the attenuance is nearly negligible over a wide energy range, from 1~meV to 1~eV, indicative of dielectric behavior. In contrast, when the polarization is parallel to the alignment direction, strong absorption occurs, especially in the midinfrared (MIR) range, primarily due to intraband dynamics, which represent the collective intratube motion of carriers~\cite{zhang2013plasmonic}. Additionally, in the near-infrared region, absorption is driven by excitonic interband transitions~\cite{HeEtAl2016NN} from metallic ($M_{11}$ peak) and semiconducting tubes ($S_{11}$ and $S_{22}$ peaks).
At higher energies, in the deep ultraviolet (DUV) range, both polarization states experience significant attenuation, which is linked to the $\pi-\pi^{*}$ gap at the M point of graphene's $k$-space~\cite{takagi2009theoretical}.

While experimental studies have provided valuable insights into the optical and excitonic characteristics of CNTs~\cite{Origin11, Origin13, Origin14, Origin15, Origin16}, understanding and modeling these properties numerically remains challenging due to interplay between collective excitations, particles, and quasiparticles. Most existing theoretical methods suffer from computational costs, convoluted calculations, and close-to-equilibrium bounded conditions. As a result, experimental optical spectra are often analyzed using combinations of Lorentzian and polynomial functions~\cite{Origin26, KatsutaniEtAl2019PR}. This approach has limitations because the baseline polynomial lacks physical significance and results can vary based on the chosen functions.

Dal~Forno \emph{et al}.~\cite{Origin} have developed a model to investigate the optical absorption of aligned single-chirality (6,5) SWCNT films. By employing group theory and pseudo-momentum conservation principles, while considering the microscopic interactions between photons, phonons ($\mathbb{A}, \mathbb{E}_1$ and $\mathbb{E}_2$), excitons ($E_{11}, E_{22}$, and $E_{12}$), and their dispersions, the model characterize seven possible absorption scattering channels.
For the case where the light polarization direction is parallel (${\rm pht}_{\parallel}$) to the CNT alignment direction, the categories are
    (1)~${\rm pht}_{\parallel}\longleftrightarrow E_{ii}$, 
    (2)~${\rm pht}_{\parallel}\pm\mathbb{A}\longleftrightarrow E_{ii}$, and
    (3)~${\rm pht}_{\parallel}\pm\mathbb{E}_1\longleftrightarrow E_{ij}$,
where $i=1, 2$ and $j=1, 2$. The first channel involves direct photon absorption or emission. Channels (2) and (3) are phonon-assisted, with $-$ (+) indicating Stokes (anti-Stokes) transitions.
Conversely, the categories of scattering for light with perpendicular polarization (${\rm pht}_{\perp}$) are
    (4)~${\rm pht}_{\perp}\longleftrightarrow E_{ij}$, 
    (5)~${\rm pht}_{\perp}\pm\mathbb{A}\longleftrightarrow E_{ij}$,
    (6)~${\rm pht}_{\perp}\pm\mathbb{E}_1 \longleftrightarrow E_{ii}$, and
    (7)~${\rm pht}_{\perp}\pm\mathbb{E}_2 \longleftrightarrow E_{ij}$.

\begin{figure}[t]
    \centering
    \includegraphics[width=10cm]{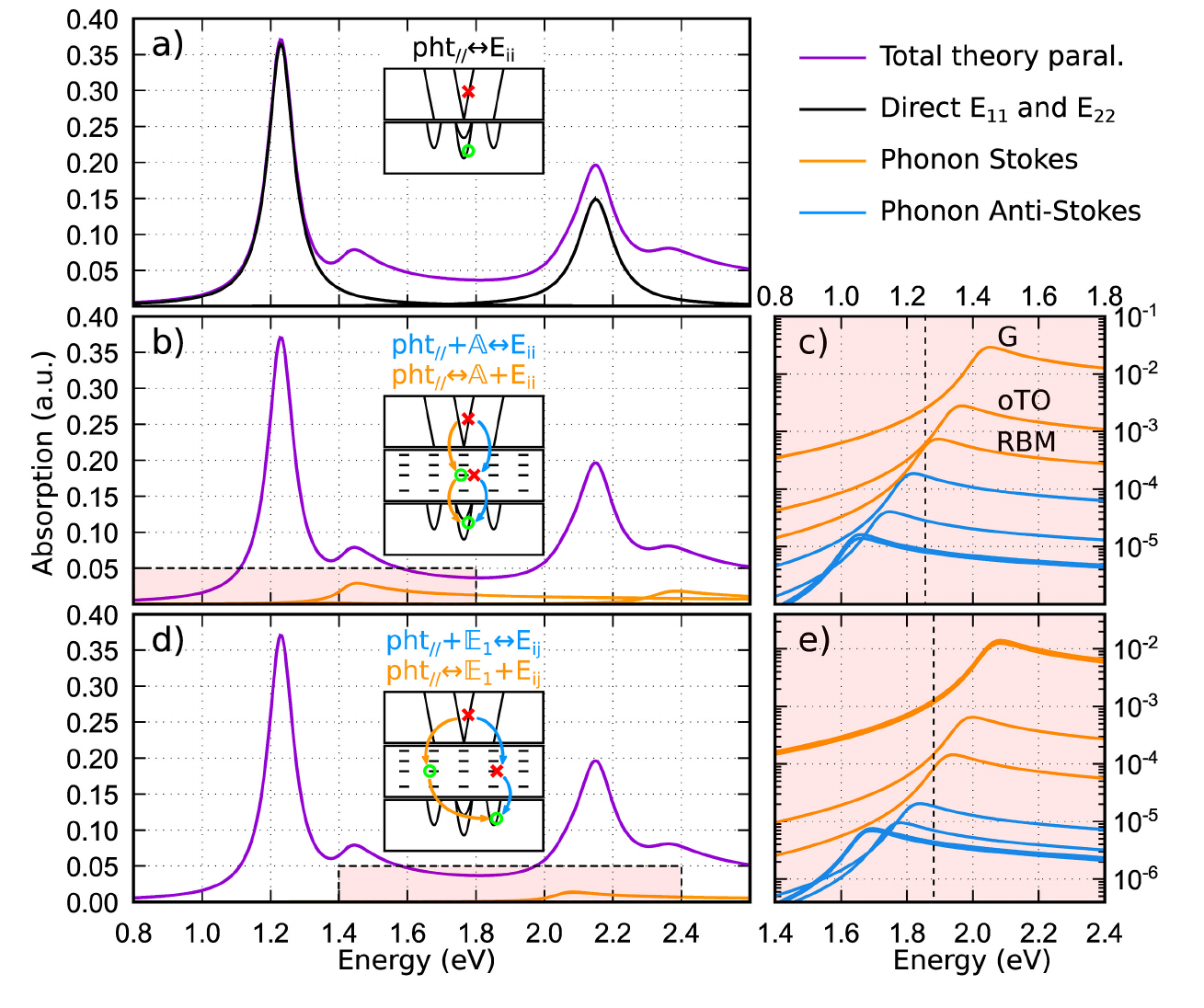}
    \caption{Absorption spectra for an aligned (6,5) CNT film calculated for an incident beam with polarization parallel to the alignment direction. (a), (b), and (d) show the different components of the spectrum corresponding to Channels (1), (2), and (3), respectively. (c) and (e) are a magnified version of the spectrum in the red-shaded rectangular area, plotted on a logarithmic scale. The vertical dashed lines represent the energy position of the lower exciton band. Reprinted from~\cite{Origin}, Copyright (2022), with permission from Elsevier.}
    \label{figure_Forno}
\end{figure}

Without loss of generality, one can focus on the parallel polarization case, corresponding to Channels (1), (2), and (3), as shown in Fig.~\ref{figure_Forno}. Specifically, Fig.~\ref{figure_Forno}(a) illustrates the dominance of the $E_{11}$ peak at 1.23~eV and the significant contribution of $E_{22}$ at 2.15~eV. Figure~\ref{figure_Forno}(b) presents the absorption contributions of $E_{11}$ and $E_{22}$ excitons assisted by $\mathbb{A}$ phonons. Additionally, the excitation dynamics of the cross-polarised exciton $E_{12}$, assisted by $\mathbb{E}_1$ phonons, are depicted in Fig.~\ref{figure_Forno}(d). The spectrum exhibits tails that stretch to higher energies. Such tails originate from the joined density of states related to transitions that involve phonon momentum transfer to higher excitonic states. Further, Figs.~\ref{figure_Forno}(c) and (e) indicate that the $G^+$ and $G^-$ phonons play a dominant role in sideband absorption, with a negligible presence of anti-Stokes scattering.

One important conclusion of the study by Dal~Forno \emph{et al}.~\cite{Origin} is that the background absorption commonly observed experimentally in CNT samples is due to phonon-assisted transitions from the semiconductor vacuum to finite-momentum continuum states of excitons. Their model successfully reproduced the experimental optical spectra of the aligned (6,5) SWCNTs with significant precision, distinguishing each scattering contribution. Therefore, this study showed that the origin of the continuum baseline observed in absorption spectra for CNTs is intrinsic, rather than due to impurities, defects, or intertube interactions.

\begin{figure}[t!]
    \centering
    \includegraphics[width=\textwidth]{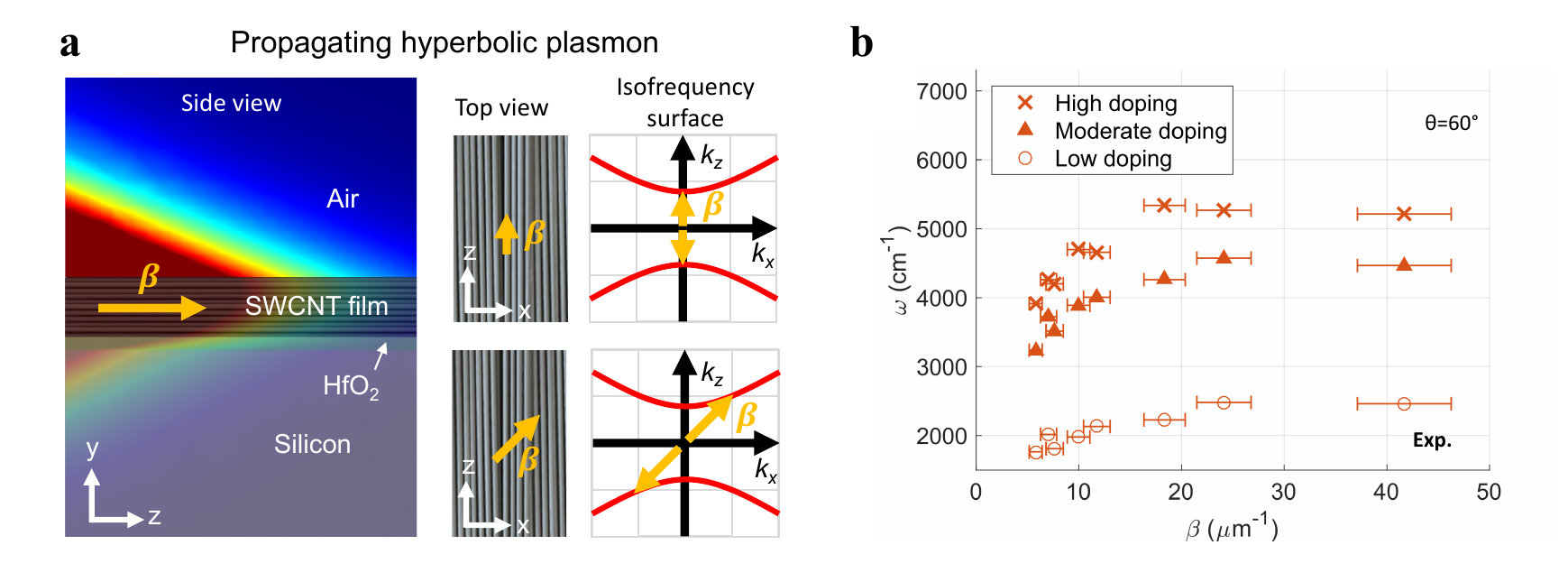}
    \caption{Hyperbolic plasmon modes (HPMs) in aligned CNT films. (a)~Left: electric field profile of a HPM propagating in a CNT film. Right: top views of in-plane HPM wavevectors along (top) and at an angle (bottom) relative to the SWCNT alignment axis, along their corresponding isofrequency contours. 
    (b)~Dispersion relations of the HPM for $\theta=60^\circ$ at various doping levels, demonstrating active tuning.
    Adapted with permission from~\cite{RobertsEtAl2019NL}. Copyright 2019 American Chemical Society.}
    \label{figure_9}
\end{figure}

Metamaterials are artificial materials, of which the electromagnetic properties are governed by the properties of subwavelength unit cells and their couplings, which allows for exceptional electromagnetic response that cannot be obtained in natural materials. Hyperbolic metamaterials, in particular, have been found to have intriguing applications, including negative refraction, sub-diffraction-limited imaging, thermal emission engineering, and sensing~\cite{PoddubnyetAl13NP,SmithetAl04JOSAB,PodolskiyNarimanov05PRB,BelovetAl05PRB,LiuEtAl2007Science,noginov2013focus,LiuShen13PRBB,RileyEtAl2017PNAS,DyachenkoEtAl2016NC,SreekanthEtAl2016NM,NaikEtAl2014PNAS}. Recently, hyperbolic properties of highly aligned SWCNT films in the MIR have been demonstrated~\cite{GaoEtAl2019ACSP,RobertsEtAl2019NL,SchocheEtAl2020JoVS,jerome2022outcoupling}. The region of frequencies where a particular film shows a hyperbolic dispersion depends on the doping level (i.e., the position of the Fermi energy) and the film thickness. To date, the hyperbolic properties of highly aligned SWCNT films have been tuned in a wide range, from $\sim$1.6\,$\upmu$m to $\sim$14\,$\upmu$m.

Roberts \emph{et al}.\ have probed the hyperbolic nature of aligned SWCNT films by exploring their bulk optical properties along patterned CNT nanoribbon's hyperbolic plasmon modes (HPMs)~\cite{RobertsEtAl2019NL}. Figure~\ref{figure_9}(a) depicts the confinement of HPMs in a CNT film and their decay into the substrate and air. Due to the in-plane hyperbolic dispersion of HPMs, the propagation of HPMs at an angle with respect to the CNT alignment direction results in large modal wavevectors $\beta$, and therefore, are promising for deep-subwavelength light confinement.
After patterning CNT films into nanoribbons to enable the direct coupling of free-space radiation and confined HPMs, Roberts \emph{et al}.\ obtained the dispersion of HPMs, as shown in Fig.~\ref{figure_9}(b). Moreover, they found that the dispersion relation is sensitive to doping (the Fermi energy), which could be used for active tuning of photonic topological transitions and epsilon-near-zero (ENZ) modes in the MIR~\cite{RobertsEtAl2019NL}. 

\begin{figure}[t!]
    \centering
    \includegraphics[width=\textwidth]{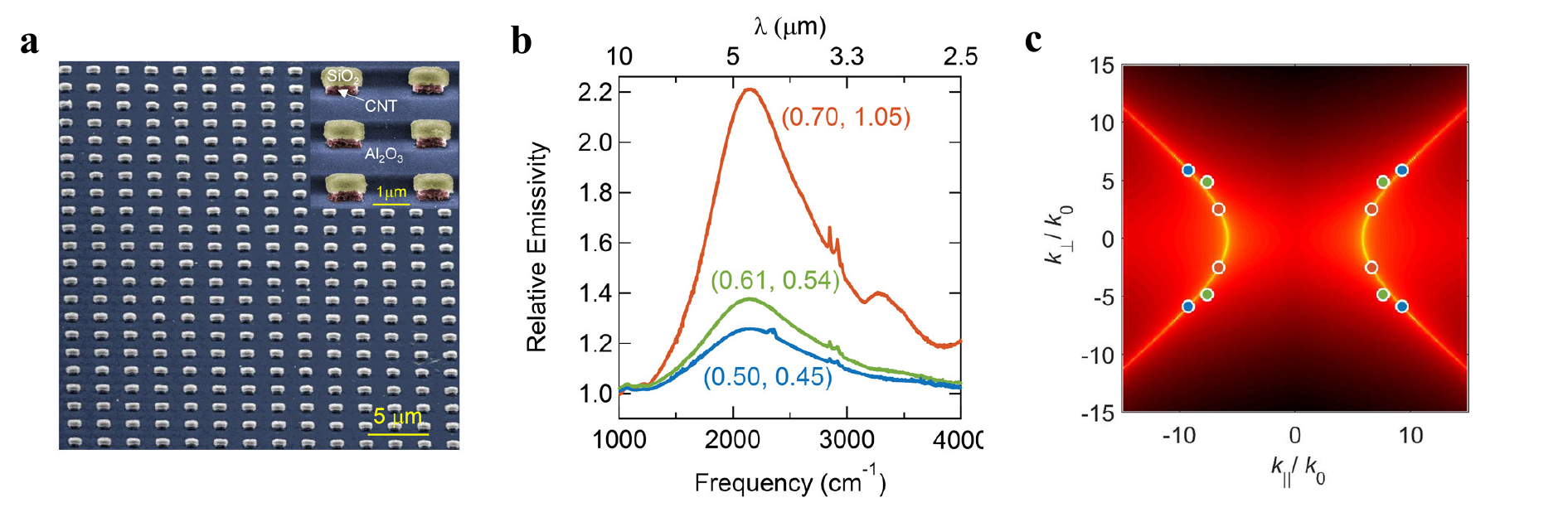}
    \caption{Thermal emission from an array of cavities made from an aligned CNT film. (a)~False-color SEM micrographs of patterned cavities. (b)~Experimental relative emissivity of patterned cavities with altered dimensions, measured in $\upmu$m. Thermal emission occurs at 2140~cm$^{-1}$ (4.7~$\upmu$m). 
    (c)~Extracted hyperbolic dispersion based on the dielectric constants at 2140~cm$^{-1}$. $k_\parallel$ is the wavevector component along the tube alignment direction ($x$-axis), $k_\perp$ is perpendicular to the alignment direction in the film plane ($y$-axis), and $k_0$ is the wavevector in vacuum. Adapted with permission from~\cite{GaoEtAl2019ACSP}. Copyright 2019 American Chemical Society.}
    \label{figure_10}
\end{figure}

Gao \textit{et al}.\ have found that, as a result of their hyperbolic properties, aligned SWCNTs provide a broadband medium of natural hyperbolic character that could be useful for MIR thermal radiation harnessing~\cite{GaoEtAl2019ACSP}. They reported spectrally selective and polarized MIR thermal emission at 700$^\circ$C. These emission spectral features arose from the Berreman modes excited near the ENZ point of the aligned CNT films. For deep-subwavelength cavities made of aligned CNTs, thermal emission was enhanced due to the propagation of high-$k$ photons in the hyperbolic medium. 
Figure~\ref{figure_10}(a) shows a representative false-color SEM image of fabricated SWCNT indefinite cavities. By simultaneously changing $L_\parallel$ and $L_{\perp}$ (which are parallel and perpendicular to the tube axis, respectively), the resonance frequency was maintained at a single frequency; see Fig.~\ref{figure_10}(b). The data points matched the isofrequency contour calculated for the SWCNT film, as shown in Fig.~\ref{figure_10}(c). The observation of resonances in deep-subwavelength-sized cavities further proved that the thermal emission was due to the high-$k$ modes, which resulted in a high photonic density of state available for thermal photons. At least a 100-fold enhancement of photonic density of states in SWCNT hyperbolic thermal emitters was observed in the smallest cavity, where a resonance had a volume of $\approx\lambda^3/700$.

\subsection{Circular polarization spectroscopy and chiroptical properties} 

\begin{figure}
    \centering
    \includegraphics[width=\textwidth]{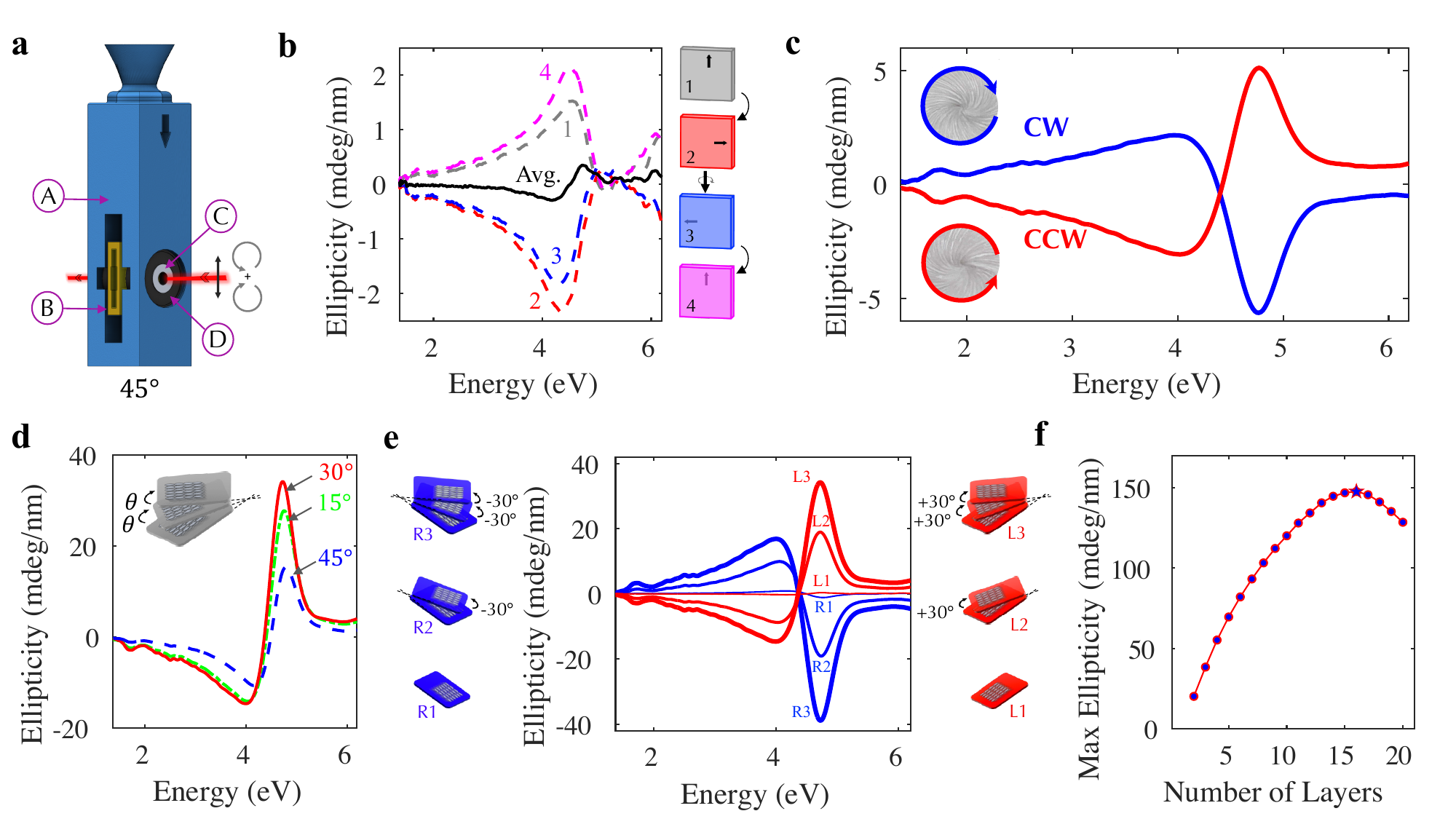}
    \caption{Characterization of the chiroptical effect in aligned and chiralized CNT assemblies. (a)~3D printed cuvette created via stereolithography, composed of (A)~the main assembly, (B)~a pocket designed for sample placement, (C)~two apertures regulating the size of the incident light beam, and (D)~two joint bearings integrated into the extruded main assembly to facilitate in-plane sample rotation. (b)~CD spectra for a highly aligned racemic (achiral) CNT film recorded in 4 configurations and their average, proving the technique's robustness. (c)~CD spectra for CNT films made using mechanical-rotation-assisted CVF with CW and CCW rotation handedness (inset: optical images of the fabricated films). (d)~CD spectra for right-hand-twisted trilayer CNT films at three different rotation angles. (e)~CD spectra for left- and right-handed CNT stacks with different numbers of layers. (f)~Calculated ellipticity as a function of number of aligned layers. Adapted from~\cite{doumani_controlled_2023}. \href{https://creativecommons.org/licenses/by/4.0/}{CC BY 4.0.}}
    \label{figure_8}
\end{figure}

When an electromagnetic wave passes through a spiral-like structure, it is subject to differences in absorption between left and right circular polarized light, resulting in circular dichrosim (CD), which is a measure of the material's helical properties. CD is usually quantified by the ellipticity ($\psi$), which is defined as $\psi \equiv \arctan\{ (E_\mathrm{l} - E_\mathrm{r})/(E_\mathrm{l} + E_\mathrm{r}) \}$, where $E_\mathrm{l}$ and $E_\mathrm{r}$ are the electric field strength of the incident left and right circular polarized light, respectively. Here, $\psi$ is usually expressed in mdeg, but it can be expressed in mdeg/nm when normalized by the sample thickness.

To isolate the genuine CD signal, CD$_\textrm{true}$, from the measured signal, $\text{CD}_\textrm{measured} = \text{CD}_\textrm{true} + 0.5\,(\text{LB}\,\text{LD}' + \text{LB}'\,\text{LD}) + \sin\alpha\,\{\text{LD}'\sin(2\theta) - \text{LD}\cos(2\theta)\}$, where $\alpha$ is the residual static birefringence of the modulator, $\theta$ is the in-plane sample rotation angle, LD is linear dichroism, and LB is linear birefringence, Doumani \textit{et al}.\ employed a four-configuration measuring protocol~\cite{doumani_controlled_2023}. They designed and manufactured a cuvette for solid-type samples, equipped with two bearings and two apertures, as shown in Fig.~\ref{figure_8}(a). This design enabled sample rotation around the optical axis by $360^{\circ}$ and out-of-plane rotation by $180^{\circ}$. By measuring the sample in four orientations, as illustrated in Fig.~\ref{figure_8}(b), and averaging the results, the authors were able to extract CD$_\textrm{true}$. 

Doumani \textit{et al}.\ created a tornado-like CNT structure by combining twisting mechanisms with alignment through orbital shaking during CVF; see Fig.~\ref{figure_8}(c). This \textit{in~situ} chirilization approach led to a strong chiroptical effect in the DUV range ($\sim$260~nm or $\sim$4.8~eV). This spectral region is linked to the M-point in graphene's $k$-space; see Fig.~\ref{figure_7_new}. The control of the shaking handedness resulted in flipped CD signals, allowing for the engineering of mirrored right- and left-handed chirality. By employing a fast shaking speed of 140~RPM, they could simultaneously induce both chiralities in a single sample during a single run. In contrast, at a slower shaking speed of 50~RPM, the ellipticity sign remained constant throughout the entire film. This observation supports the idea that chiroptical responses resulted from light propagating along the axis of the engineered 3D helical structure.

\begin{table}[width=.9\linewidth,cols=5,pos=b]
    \centering
    \caption{\bf \centering Comparison of different chiral platforms in the deep ultraviolet spectral range~\cite{doumani_controlled_2023}.}
    \begin{tabular}{|c|c|c|c|}
    \hline
    Platform   & \makecell{Wavelength (nm)} & \makecell{Ellipticity (mdeg/nm)} & Reference \\
    \hline\hline
    \makecell{TiO$_2$ metamaterial}   & $\sim270$ & $\sim17$ & Ref.\,\cite{SarkarEtAl2019NL}\\
    \hline
    \makecell{Colloidal Al} & $\sim280$ &  $\sim4$ & Ref.\,\cite{HuangEtAl2022OE}\\
    \hline
    \makecell{Mg metamaterial}  & $\sim275$ &  $\sim16$ & Ref.\,\cite{JeongEtAl2016CC} \\
    \hline
    \makecell{Al metamaterial} & $\sim250$ &  $\sim8.6$ & Ref.\,\cite{LeiteEtAl2022NL}\\
    \hline
    \makecell{Bilayer graphene} & $\sim280$ & $\sim3$ & Ref.\,\cite{KimEtAl2016NN, ChaeEtAl2011NL}\\
    \hline
    \makecell{(MBA)$_2$PbI$_4$} & $\sim220$& $\sim0.2$ & Ref.\,\cite{LuEtAl2019SA}\\
    \hline
    \makecell{Trilayer CNT} & $\sim260$ & $\sim40$ & Ref.\,\cite{doumani_controlled_2023}\\
    \hline
    \end{tabular}
      \label{Benchm}
\end{table}

Doumani \textit{et al}.\ also utilized an alternative, \textit{ex~situ} approach centered on 3D-twist stacking to further explore the structural chirality of CNTs assemblies (Sec.~\ref{CD_fab}). The strength of the induced ellipticity was defined by the angle at which the layers were stacked. They discovered a record-high DUV ellipticity of $40 \pm 1$~mdeg/nm at 260~nm, which could be achieved with a $30^{\circ}$ twist for three layers; see Fig.~\ref{figure_8}(d). The dissymmetry factor $g$ was calculated to be $\sim$0.07 by taking the ratio of CD to the average absorption of left and right circularly polarized light. This $g$ value is larger than that of chiral molecules two orders of magnitude and comparable to other materials; see Table~\ref{Benchm} for more details.

Figure~\ref{figure_8}(e) show that, by controlling both the twist-angle handedness and the number of stacked layers, it is possible to engineer the chirality handedness and tune the ellipticity power. Furthermore, Doumani \textit{et al}.\ demonstrated the ability to switch CD on and off by alternating between angle handedness for these stacks. For instance, in ABAB-type stacking, an odd number of layers, such as A and ABA, resulted in nearly negligible CD, while even numbers such as AB and ABAB exhibited finite ellipticity.
The experimental validation involved ellipsometry spectroscopy, from which the 16 Müller matrix elements were obtained, with which a differential decomposition approach was applied. Additionally, they bolstered their findings with numerical simulations based on the transfer matrix formalism. These simulations predicted that with a 16-layer stack, as highlighted in Fig.~\ref{figure_8}(f), an ellipticity as large as 150~mdeg/nm should be reached, equivalent to a $g$ factor of 0.22.

\section{Device Applications of Macroscopically Ordered CNT Assemblies}

The advancements made in the fabrication and basic studies of macroscopically ordered CNTs, described in the previous sections, have stimulated a multitude of new ideas for applications~\cite{zubair2016carbon,komatsu2021macroscopic,liu2020aligned,evanoff2012towards}. In this section, we discuss specific potential device applications of highly aligned nanotube films based on their macroscopically anisotropic optoelectronic properties.

\begin{figure}[t!]
    \centering
    \includegraphics[width=0.95\textwidth]{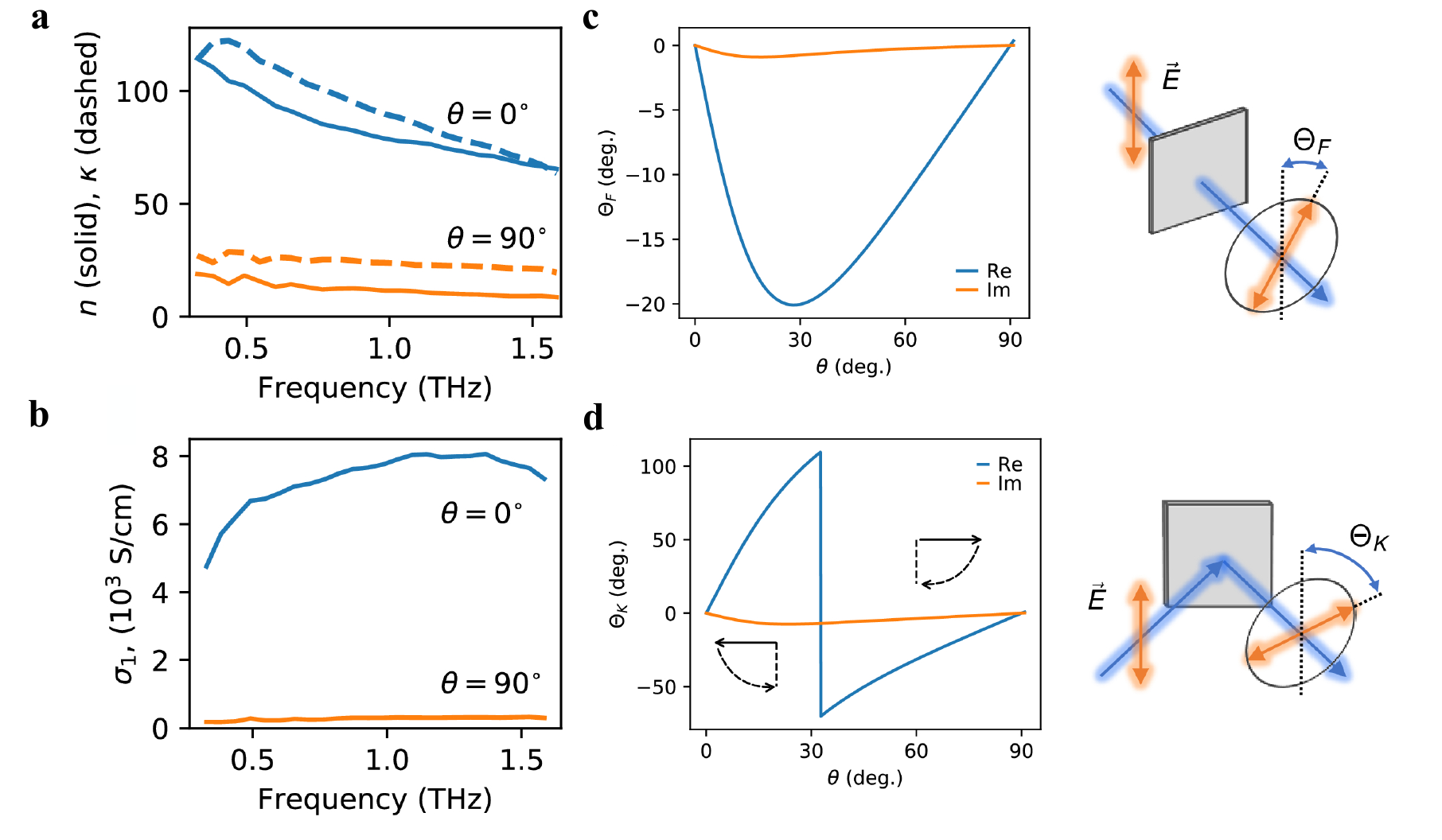}
    \caption{Optical anisotropy and polarization rotation in the THz range in a highly aligned CNT film. (a)~Real (solid line) and imaginary (dashed line) parts of the polarized refractive index. (b)~Real part of the optical conductivity. (c)~Complex angle of polarization rotation upon transmission and (d) reflection from the CNT film. Schematics on the right show the definition of the polarization rotation angles. Adapted with permission from~\cite{BaydinetAl21Optica}. \copyright\,The Optical Society.}
    \label{figure_12}
\end{figure}

\subsection{THz wave manipulation and modulation}

In recent years, the terahertz (THz) range has gained considerable attention due to its potential applications in optics, sensing, and communication~\cite{Tonouchi2007}. However, traditional optical components in this spectral range, such as polarizers, waveplates, and filters, exhibit various limitations, including brittleness, lack of flexibility, low extinction ratios, and bulky designs~\cite{yeh1978new, wiesauer2013recent, masson2006terahertz}.
This has created a growing demand for innovative optical elements and materials specifically designed to manipulate THz radiation. While some progress has been made in developing components with polarization rotation capabilities, advanced elements such as metamaterials~\cite{Kan2015, Deg2022} still face challenges related to scalability and production complexity.

In a recent study, using THz time-domain spectroscopy (TDS), Baydin \emph{et al}.~\cite{BaydinetAl21Optica} reported record-breaking polarization rotation effects in an aligned CNT film, reaching $\sim$20$^{\circ}$ during single-pass transmission and $\sim$110$^{\circ}$ upon a single reflection. These phenomena were attributed to the anisotropic optical properties achievable with aligned CNT films prepared using CVF, resulting in large differences in the bi-axial refractive indices. During transmission, as shown in Figs.~\ref{figure_12}(a) and (b), the electric field parallel to the CNT alignment direction ($E_\parallel$) experiences larger attenuation than the perpendicular field ($E_\perp$), as described in Sec.~\ref{spectro}, and $E_\parallel$ is reflected to a greater extent than $E_\perp$. The angles of rotation, $\Theta_F$ (Faraday) for transmission and $\Theta_k$ (Kerr) for reflection, depended on the angle ($\theta$) between the incident light polarization and the CNT alignment direction; see Figs.~\ref{figure_12}(c) and (d). However, it is essential to note that in cases of pronounced attenuation along one direction, such as with wire polarizers, this phenomenon will not be observed.
As a result, in their work, Baydin \emph{et al}.\ demonstrated a robust, broadband, thin, flexible, tunable, and simple-to-fabricate THz polarization component with pronounced birefringence and optical polarization rotation.

\subsection{Solar cells}

\begin{figure}[t!]
    \centering
    \includegraphics[width=0.95\textwidth]{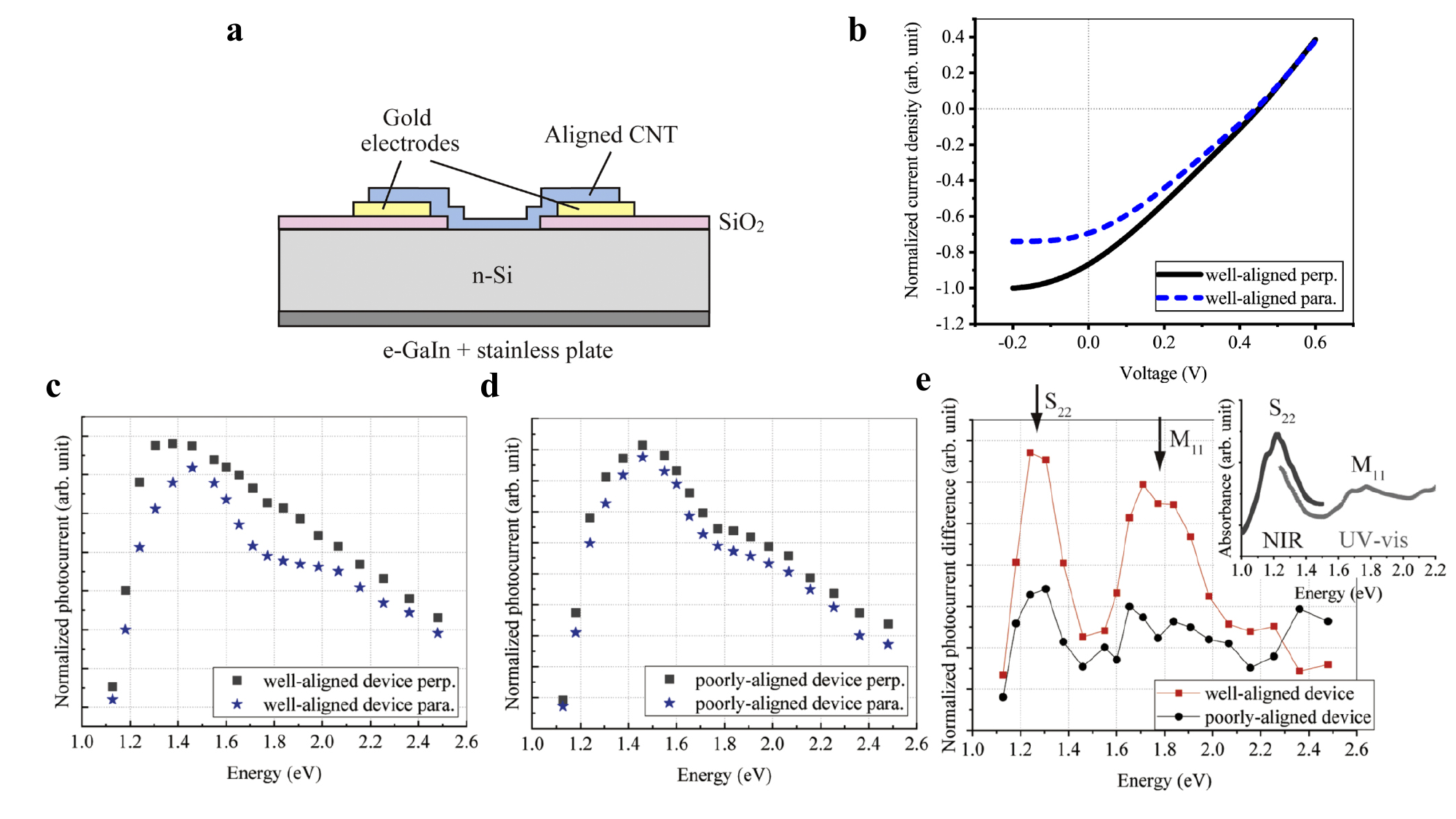}
    \caption{Aligned CNT films integrated into solar cell technologies. (a)~Schematic illustration of an ordered CNT/Si heterojunction solar cell. (b)~I–V curve of an aligned CNT device illuminated by linearly polarized light parallel and perpendicular to the direction of alignment of nanotubes. Normalized photocurrent in the parallel and perpendicular polarization cases of (c)~a higly aligned CNT-based Si-CNT solar cell and (d)~a poorly aligned device. (e)~Difference (photocurrent in the perpendicular situation minus that in the parallel situation) in the normalized photocurrent of each solar cell. Inset: absorption spectrum of the original SWCNT solution. Adapted from~\cite{SolarCell}. \copyright\,The Japan Society of Applied Physics.  Reproduced by permission of IOP Publishing. All rights reserved.}
    \label{figure_13}
\end{figure}

Use of CNTs in solar cell technology~\cite{SolarCell1, SolarCell2, SolarCell3} has shown promise in enhancing efficiency~\cite{SolarCell4} and cost-effectiveness~\cite{SolarCell5}. However, researchers are divided on the nature of the Si-CNT junction, with some proposing that it functions as a $p$-$n$ junction~\cite{SolarCell6} and others arguing that it is a Schottky junction~\cite{SolarCell7}. In addition, research by Jia \emph{et al}.\ has suggested that the CNTs act more similarly to metallic materials~\cite{SolarCell8}. They found that inserting an insulating layer at the CNT-Si interface improved the performance of the solar cells. The use of randomly oriented CNT films and various chiralities has posed challenges in elucidating the dominant mechanisms.

In a recent study~\cite{SolarCell}, Nakamura \emph{et al}.\ used aligned CNT films made by CVF in Si-CNT solar cells to explore the photoconversion mechanism; see Fig.~\ref{figure_13}(a). They investigated how aligned CNT films affect light absorption based on incident light polarization angles, largely influencing film absorbance. When the polarization was parallel to the CNT alignment direction, the short-circuit photocurrent decreased by approximately $25\%$; see Fig.~\ref{figure_13}(b). Based on this observation, they concluded that electron-hole pair generation in the CNT layer does not directly lead to power generation. This work also suggested that CNT-Si devices are Schottky junctions, not $p$–$n$ junctions, a key distinction for future development.

Figures~\ref{figure_13}(c) and (d) show the photocurrent's wavelength dependence under linearly polarized light, normalized by the incident light intensity. In well-aligned CNT devices, photocurrent reduction occurred, primarily at around 1.3 and 1.8~eV, when the light polarization aligned with the CNT alignment direction, whereas poorly aligned devices exhibited a consistent photocurrent decrease across all wavelengths. Figure~\ref{figure_13}(e) displays the photocurrent difference between parallel and perpendicular orientations, with peaks corresponding to the expected CNT interband transitions. This reduction in photocurrent can be linked to CNT-absorbed light energy, even in the presence of semiconducting CNTs. Therefore, it can be again concluded that the fabricated device does not function as a conventional $p$–$n$ junction, indicating a limited contribution of light absorbed by CNTs to the generated power.

\section{Conclusion and Prospects}
Recent advances in fabricating highly aligned CNT films have enriched our understanding of alignment mechanisms, paving the way for consistent and scalable production, supported by refined morphological characterization, machine learning, and automation. Studies using linearly polarized spectroscopy have deepened our insight into interband transitions and the hyperbolic behaviors exhibited by these highly anisotropic materials. Novel methods like twist stacking and mechanical-shaking-assisted CVF have achieved record-high ellipticity, facilitating the creation of versatile, wafer-scale, chiral thin films with customizable chiroptical properties. Ordered CNT architectures have proven to be effective in applications, including THz polarization rotation enhancing silicon-based solar cell junctions. Moreover, these ordered CNTs offer versatility in radial flexible assemblies, promising diverse applications, such as thermal management, EUV photomasks, radial polarizers, secured communication platforms, and bio-sensors.

These advancements underscore the pivotal role of alignment and structural-ordering of nanotubes, providing a foundation for future research endeavors. This exploration into novel geometries encourages ongoing dedication to improving fabrication methods and discovering new CNT assemblies with potential applications. Additionally, the potential to extend current CNT ordering techniques to a broader range of materials, spanning 0D, 1D, and 2D structures, and creating composites with diverse materials reveal promising avenues for exploring captivating physics and delving into exciting possibilities in further investigations.

\appendix

\bibliography{AB-refs,SS-refs,GR1refs,SY-refs,GR2ref,keshav,JD-refs,Jun}

\end{document}